\documentclass{article}

\pdfoutput=1
\usepackage{fullpage}
\usepackage[utf8]{inputenc}
\usepackage{graphicx}
\usepackage{amssymb}
\usepackage{amsmath}
\usepackage{amsfonts}
\usepackage{multirow}
\usepackage{tabularx,colortbl}
\usepackage{color}
\usepackage[pdftex]{hyperref}
\usepackage[english]{babel}
\usepackage{subfigure}
\usepackage{algorithm}
\usepackage{algorithmic}
\usepackage{lineno}
\usepackage[sort&compress,numbers]{natbib}
\usepackage{authblk}

\definecolor{lgray}{rgb}{0.92,0.92,0.92}

\makeindex

\begin{document}

\title{Toward a multilevel representation of protein molecules: comparative approaches to the aggregation/folding propensity problem}

\author[1]{Lorenzo Livi\thanks{llivi@scs.ryerson.ca}\thanks{Corresponding author}}
\author[2]{Alessandro Giuliani\thanks{alessandro.giuliani@iss.it}}
\author[3]{Antonello Rizzi\thanks{antonello.rizzi@uniroma1.it}}
\affil[1]{Dept. of Computer Science, Ryerson University, 350 Victoria Street, Toronto, ON M5B 2K3, Canada}
\affil[2]{Dept. of Environment and Health, Istituto Superiore di Sanit\`{a}, Viale Regina Elena 299, 00161 Rome, Italy}
\affil[2]{Dept. of Information Engineering, Electronics, and Telecommunications, SAPIENZA University of Rome, Via Eudossiana 18, 00184 Rome, Italy}
\renewcommand\Authands{, and }
\providecommand{\keywords}[1]{\textbf{\textit{Index terms---}} #1}

\maketitle

\begin{abstract}
This paper builds upon the fundamental work of \citet{niwa2009}, which provides the unique possibility to analyze the relative aggregation/folding propensity of the elements of the entire Escherichia coli (E. coli) proteome in a cell-free standardized microenvironment.
The hardness of the problem comes from the superposition between the driving forces of intra- and inter-molecule interactions and it is mirrored by the evidences of shift from folding to aggregation phenotypes by single-point mutations \cite{doi:10.1021/ja1116233}.
Here we apply several state-of-the-art classification methods coming from the field of structural pattern recognition, with the aim to compare different representations of the same proteins gathered from the Niwa et al. data base; such representations include sequences and labeled (contact) graphs enriched with chemico-physical attributes.
By this comparison, we are able to identify also some interesting general properties of proteins. Notably, (i) we suggest a threshold around 250 residues discriminating ``easily foldable'' from ``hardly foldable'' molecules consistent with other independent experiments, and (ii) we highlight the relevance of contact graph spectra for folding behavior discrimination and characterization of the E. coli solubility data.
The soundness of the experimental results presented in this paper is proved by the statistically relevant relationships discovered among the chemico-physical description of proteins and the developed cost matrix of substitution used in the various discrimination systems.\\
\end{abstract}

\section{Introduction}

The Anfinsen's paradigm states that the native structure of a protein is encoded in its primary structure \cite{Tanuichi}.
This paradigm comes from in vitro unfolding/refolding experiments. While certainly Anfinsen’s paradigm is valid in principle, it is severely hampered in the actual cell environment.
As a matter of fact, many protein species require the intervention of a variety of ``chaperone'' proteins in order to undergo a correct folding process, ending up into the correct soluble native structure.
Protein folding competes with intermolecular aggregation that is driven by very similar forces but ends up into a very different final state: instead of producing soluble and independent molecules, aggregation provokes the formation of insoluble bodies made by many molecules \cite{Agostini2012237}.
The balance between ``within-molecule-interaction'' and ``between-molecules-interactions'' shifts the equilibrium towards correct folding (prevalence of within-molecule-interactions, soluble systems) or aggregation (prevalence of between-molecules-interactions, insoluble systems).
This balance can be so subtle that in some cases a single point mutation \cite{doi:10.1021/ja1116233} is able to provoke a dramatic shift of this equilibrium. It is important to stress that, despite the presence of chaperones, a certain level of aggregation is present in the cells for any protein system, moreover the cell chemico-physical microenvironment (pH, crowding, etc.) has a crucial importance in determining the folding/aggregation propensity of different protein systems.
The existence of several diseases provoked by misfolding and consequent aggregation of protein molecules \cite{doi:10.1146/annurev.biochem.75.101304.123901} gives the elucidation of folding/aggregation balance a great applicative importance.
In their 2009 paper appeared on PNAS, Niwa and colleagues \cite{niwa2009} produced the most precise and pure data base upon which test different explanatory models of folding/aggregation determinants.
At odds with other studies, the authors generated a fair test bench in which the complete ORF library (Open Reading Frames; they correspond to the structural genes of a given genome, i.e., the set of the sequences that are translated into protein molecules) of E. coli was translated in a pure, cell-free and thus chaperone-free, system under the same conditions.
This allows to get data on around 70\% of the E. coli proteins and can be considered as the best experimental framework on which test the intrinsic propensity of a given sequence to go toward the aggregation or correct folding pole.
This is a very crucial point that deserves further explanation. In general, solubility cannot be considered as an intrinsic property of the solute molecule, given that it depends from the solvent and chemico-physical microenvironment features, like temperature, presence of other molecules, pH, ionic strength etc.
The Niwa et al. data set, in which microenvironment is kept invariant, allows to concentrate solely on solute (protein molecule) properties, so making it possible to devise a meaningful pattern recognition strategy.
As a matter of fact, any protein whose 3D structure is known is amenable to be solubilized in some way, but this fact does not imply that can be solubilized by the Niwa et al. minimalistic recipe.
Strictly speaking, \citet{niwa2009} observed the relative solubility of different protein species; only as a second step not correctly folded molecules precipitate and aggregate. In a recent paper, \citet{Agostini2012237} clearly demonstrated that the solubility degree in the Niwa et al. data base negatively correlates with the aggregation propensity, when the aggregation is estimated from the folded state.
This implies that we can safely consider the solubility as a measure of the relative stability of the folded and aggregated states \cite{Agostini2012237}.

The authors observed a bi-modal distribution as for the relative solubility of the 3173 protein species assayed \cite{niwa2009} consistent with the presence of two classes of molecules: proteins not necessitating any chaperone to reach the correct stable folding and proteins that need chaperone intervention in order to avoid aggregation.
This bi-modal character is far to be perfect and there are large areas in between the two peaks confirming the fuzzy boundaries between the two folding behaviors.
The authors describe some general properties showing statistically significant trends between the two groups of soluble and insoluble proteins but they were not able to predict folding/aggregation propensity on a single molecule scale using both sequence and final 3D structure based algorithms.
Thus the authors conclude that proteins more prone to aggregate are bigger than more soluble ones, have a lower isoelectric point, and a higher content of negatively charged residues \cite{niwa2009}.
On the structural side, all-$\beta$ class was more frequent in the low-solubility population like some specific SCOP folds, such as c-94 or c-67. However, the authors were not able to go beyond a population-based statistical description.
On the other hand, \citet{Agostini2012237} obtained a good prediction of actual relative solubility of proteins by a Support Vector Machine (SVM) based approach as applied to chemico-physical representations of sequences, hence giving a proof of concept of the possibility to express solubility/aggregation balance in terms of chemico-physical properties of amino acid residues.

Instead of having as main goal the one of obtaining the best correlation between continuous solubility and chemico-physical description of sequences, in this paper we compared alternative representations of the E. coli proteins as for their ability to discriminate between spontaneous foldable and chaperone-needing proteins.
We analyzed the Niwa et al. dataset by considering pattern recognition methods operating on both sequence and graph based pattern representations.
Departing from the more conventional vector-based representation of patterns implies the adoption data-driven inference mechanisms that able to operate in the so-called non-geometric input spaces \cite{si_asoc_grc}. On the other hand, such types of representations offer also more interpretable information that could be exploited to infer knowledge from the data-driven process itself.
Experimental results show the possibility to reach good accuracy in the prediction of folding/aggregation propensity of single protein systems.
This constitutes a neat progress with respect to the statistical evidence produced by the authors \cite{niwa2009}, which allows us to sketch some general hypotheses on folding process based on some features of the adopted modeling and computational approaches.
Comparing the biophysical/biochemical premises of successful vs. unsuccessful predictions, we were able to single out some important general properties of protein architectural and dynamical features.
Notably, (i) we suggest a threshold around 250 residues discriminating ``easily foldable'' from ``hardly foldable'' molecules consistent with other independent experiments, and (ii) we highlight the relevance of contact graph spectra for folding behavior discrimination and characterization of the E. coli solubility data.
The contact graph spectral description, which we have found to be suitable for this specific problem, can be generalized to other network classification problems and considering different structure-function relationships.

The remainder of the paper is structured as follows. In Section \ref{sec:material} we first present the considered datasets (Section \ref{sec:dataset}) corresponding to alternative protein molecule representations, which we elaborated starting from the original data of \citet{niwa2009}. Then we introduce the methodological basis underlying the particular classification systems used for processing such data (Section \ref{sec:class}).
In Section \ref{sec:complexity}, we describe the theoretical frameworks that we used to experimentally evaluate the structural complexity of the graphs representing the E. coli proteins.
In Section \ref{sec:results}, we present and discuss the experimental results, which include: (i) a statistical analysis demonstrating the consistency and soundness of our results, (ii) the detailed test set classification accuracy percentages obtained on the various datasets, and finally (iii) a preliminary analysis focusing on the study of the structural complexity of the developed protein graph representations.
Finally, Section \ref{sec:conclusions} concludes the paper pointing at future directions.

\section{Materials and methods}
\label{sec:material}

\subsection{Elaboration of datasets}
\label{sec:dataset}

\begin{figure}[ht!]
 \centering
 \includegraphics[viewport=0 0 341 243,scale=1,keepaspectratio=true]{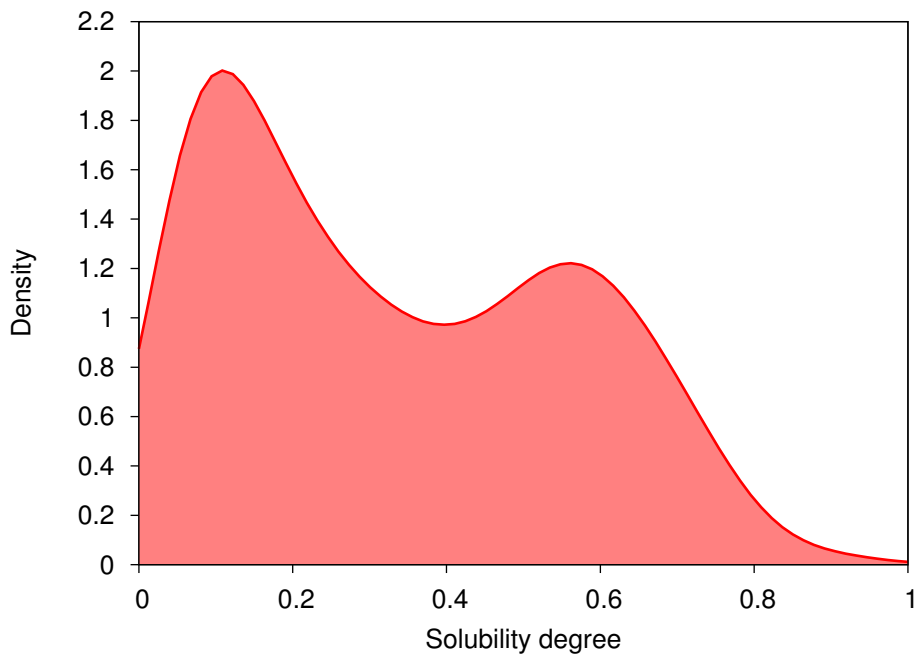}
 \caption{Density of the normalized solubility degree of proteins.}
 \label{fig:solubility_density}
\end{figure}

Niwa and colleagues \cite{niwa2009} demonstrated a bi-modal distribution of solubility with a class of largely insoluble and a class of very soluble proteins.
In Figure \ref{fig:solubility_density}, we report the plot showing the protein solubility density according to the normalized solubility degree; each protein solubility value has been normalized by considering the maximum solubility degree in the dataset.
This unbalanced bi-modal character of the solubility makes difficult the selection of a universally valid threshold partitioning proteins into soluble and not soluble hard categories.
In fact, considering $[0, 0.3]$ and $[0.7, 1]$ as the two intervals characterizing the insoluble and soluble proteins, the original dataset would split into 1631 insoluble and 180 soluble proteins, respectively.
The original data is provided in terms of a sequence-based representation of the proteins, which basically includes only the symbolic information of the residues (i.e., a character that identifies an amino acid).

The experiments that we conducted have been organized by considering two different sub-samplings of the original dataset. The first sub-sampling considers a perfectly balanced dataset made of the 100 proteins with the highest solubility degree and the 100 proteins with the lowest solubility degree.
Such a dataset, which is evidently a ``simplification'' of the original problem, has been analyzed in our previous work \cite{grapsec_ijcnn_2013}, where we demonstrated the possibility to achieve a very good classification accuracy.
The second dataset sub-sampling, which we elaborate here in this paper, determines the insoluble and soluble classes by taking the proteins falling in the $[0, 0.3]$ and $[0.7, 1]$ ranges, respectively.
The solubility degree ranges have been considered of the same interval length, by placing such intervals on the extremes of the normalized solubility degree range.
This dataset has been split into a training set and a test set.
The training set contains 180 proteins, 70 of which belong to the soluble class and the remaining 110 to the insoluble class. The test set is considerably larger, since in fact it contains 1631 proteins, of which 1521 are insoluble proteins.
Consequently, this dataset (in the following denoted as DS-1811) is largely unbalanced, especially for what concerns the test set.
The training set, although it is much smaller than the test set, it is conceived to ``cover'' the considered data instance as much as possible with \textit{characterizing} proteins, that is, with proteins that suitably represent the soluble or insoluble prototypical patterns. Each protein in DS-1811 is represented by a sequence of symbols (that is characters) associated with amino acids.

As mentioned before, our first goal is to show the possibility to achieve a good predictive ability on the sole basis of pure sequence symbolic information.
Successively, we project the solution -- in terms of substitution matrix between amino acid residues and sequence matching techniques -- on a chemico-physical space, in which the residues are encoded by the principal component scores derived by a large collection of well-known chemico-physical descriptors \cite{apdbase}.

The sequence dissimilarity measure between proteins is assessed by means of a Levenshtein-type metric.
The Levenshtein distance is a string metric that calculates the distance among two strings as the minimum cost of alignment \cite{Deza.Deza2009EncyclopediaofDistances}.
The alignment procedure is a computation that operates globally by means of the so-called edit operations of insertion, deletion, and substitution, quantifying the cost involved in transforming a string into the other.
The algorithm is usually implemented in the well-known dynamic programming framework, resulting thus in a quadratic computational complexity.
The Levenshtein distance has been originally conceived to process sequences of symbols (i.e. strings). However, it can be generalized to sequences of general objects for which it is possible to define a suitable distance measure in the domain of those objects \cite{t2vsdiss__ifsanafips2013,t2apdiss__ifsanafips2013}.
In the general case, the edit costs used to calculate the overall minimum-cost of alignment among sequences are obtained by evaluating such a distance between the objects of the sequences.
It is worth noting that the Levenshtein distance, at odds with other metrics used in protein sequences comparison, is not limited to homologue comparisons.

Successively, we tried to solve the prediction task on the structural side, by representing the proteins as graphs.
At odds with Niwa et al., we made use of a topological enriched representation in which the vertices of a graph are labeled by the three main chemico-physical components coming from a collection of amino acid chemico-physical features (described in detail in Sec. \ref{sec:cca}).
The prediction-by-structure could only be achieved for proteins whose 3D structure is known and present in the PDB database \cite{pdb}.
As a consequence, we were able to construct a dataset made by the 454 contact graphs that were available from PDB files.
We will refer to such dataset of labeled graphs as DS-G-454.

A contact graph is constructed by considering as vertices the centres of mass of each residue (i.e., their alpha carbon atoms). Edges are generated by connecting residues at a distance in between 4 and 8 \AA{}; this choice was dictated by the need to eliminate trivial contacts due to neighboring residues along the sequence while focusing on the elective contacts characterizing the native structure \cite{doi:10.1021/cr3002356}.
Each graph representing a (folded) protein is enriched by considering suitable attributes for both vertices and edges: such types of graphs are termed labeled (attributed) graphs in the pattern recognition literature \cite{gm_survey,gralg_2012,odse,odse2_ijcnn_2013,foggia2012graph}.
The vertices of the graphs are characterized with the three aforementioned chemico-physical components.
Edges are labeled by using the Euclidean distance among the residues in the 3D space. Such a characterization allows us to include in the graph representation both topological and chemico-physical information regarding proteins.
DS-G-454 is constituted by 377 insoluble and 77 soluble proteins; 27 soluble and 132 insoluble proteins have been randomly considered for the test set, while the remaining proteins are used for training.
It is worth noting that, although even in this data set we have a prevalence of insoluble proteins, the obvious fact that in order to get a reliable 3D structure from X-ray or NMR, the molecule must be solubilized. The point is that here we are not studying ``solubility as such'', which in turn is a devoid of chemico-physical sense concept, but the solubility of the molecule in the Niwa et al. specific experimental conditions, which are different from those adopted (on a case-by-case basis) for structural analysis.

By considering only the proteins in DS-G-454 represented as graphs, we have generated another dataset instance containing the related 454 sequences -- thus it is a subset of DS-1811.
We denoted this dataset DS-454. Such a dataset has been considered with the aim of evaluating the effect of sub-sampling DS-1811 from the classification performance viewpoint. DS-454 has been divided into training and test set by considering the same split percentages used for DS-G-454.

The last dataset that we considered is obtained as a direct elaboration of DS-G-454.
Notably, each graph has been mapped into a sequence of vertex attributes according to the procedure of graph serialization that we explain in the following.
By elaborating the whole dataset of graphs, we obtained another dataset of 454 sequences of three-dimensional, real-valued vectors (i.e., the three aforementioned components). We denoted such a dataset DS-S-454.
A sequence of vertices related to a graph is obtained by using the so-called seriation algorithm discussed in Ref. \cite{livi+delvescovo+rizzi_seriation+gradis,seriation+gradis_lncs_2012}.
A graph $G=(\mathcal{V}, \mathcal{E}), |\mathcal{V}|=n, $ where $\mathcal{V}$ is the set of vertices and $\mathcal{E}$ is the set of edges, can be represented according to its transition matrix, $\mathbf{T}^{n\times n}$, which is given as $\mathbf{T}=\mathbf{D}^{-1}\mathbf{A}$. The matrix $\mathbf{D}$ is diagonal and contains the vertex degrees (i.e., number of incident edges), while $\mathbf{A}$ is the adjacency (i.e., the contact) matrix of $G$.
To include the information about the edges in $\mathbf{T}$, we consider the degree of a vertex to be weighted by the quantity stored in the edge attributes (i.e., the atomic distances). The seriation of the graph is then implemented by analyzing the first eigenvector of $\mathbf{T}$ (i.e., the eigenvector associated to the largest, in magnitude, eigenvalue, i.e., 1 in this case), which corresponds to the stationary distribution, $\pi$, of the Markovian random walk on $G$ \cite{Lovasz1996}.
$\mathbf{T}$ may not be symmetric, hence it may not allow a spectral decomposition. However, $\mathbf{T}$ can be symmetrized by computing $\mathbf{D}^{-1/2}\mathbf{T}\mathbf{D}^{-1/2}$, with which shares the same eigenvectors.
Interestingly, the first eigenvector of the contact matrix is known to provide a consistent description of proteins \cite{PROT:PROT22113}, and so is the principal eigenvector of the transition matrix, although it should be interpreted in the random walk setting.
This further sequence representation of proteins allows us to deal with the recognition problem on the ``less complex domain'' of sequences, while at the same time it includes the topological information of the graph representations as a byproduct of the seriation. Training and test set splits are compatible with those of DS-G-454.

To summarize, in this study we considered four datasets related to the same initial set of proteins, namely DS-1811, DS-454, DS-G-454, and DS-S-454.
DS-1811 and DS-454 contain sequences of characters denoting the usual amino acid identifiers.
DS-G-454 is a dataset of labeled graphs describing the proteins according to their 3D native structure, including also suitable vertex--edge auxiliary information.
Finally, DS-S-454 is a direct elaboration of DS-G-454, containing sequences of three-dimensional, real-valued vectors, which are a characterization of the graphs according to the topological information provided by the corresponding transition matrices.

\subsection{The considered pattern recognition systems}
\label{sec:class}

Each of the aforementioned datasets has been processed with an ad hoc similarity-based classification system. Design of dissimilarity and similarity measures is of utmost importance in pattern recognition systems, since they effectively induce the underlying algebraic structure of the input space \cite{gm_survey}.
Kernel methods are an important class of pattern recognition systems, which are based on the definition of a (positive semi-definite) kernel function, $k(\cdot, \cdot)$, which is used to calculate the input pattern similarity \cite{schoelkopf+smola2002}.
Assuming to deal with $\mathbb{R}^d$ data, it is well-known that distances and similarity among vectors are related by the inner product $\langle\cdot, \cdot\rangle$ construct:
\begin{equation}
d(x, y) = \sqrt{\langle x, x\rangle + \langle y, y\rangle - 2\langle x, y \rangle}.
\end{equation}

Let $K_{ij}=\langle x_i, x_j\rangle$ be the $n\times n$ positive definite similarity matrix (also called Gram matrix), containing the pairwise inner products among the $n$ input patterns. Let $D_{ij}=d(x_i, x_j)$ be the matrix containing the pairwise Euclidean distances. It is possible to obtain $\mathbf{K}$ by computing
\begin{equation}
\mathbf{K} = -\frac{1}{2}\mathbf{C}\mathbf{D}^2\mathbf{C},
\end{equation}
where $\mathbf{C}=\mathbf{I}-\mathbf{1}/n$ is called centering matrix and $\mathbf{D}^2$ contains the squared distances \cite{Duin2012826,si_asoc_grc}.
Since the evaluation of a positive definite kernel function $k(\cdot, \cdot)$ corresponds to the evaluation of an inner
product on an \textit{induced} high-dimensional Hilbert space (Mercer’s theorem), we can determine the similarity matrix as $K_{ij}=k(x_i, x_j)$.
Therefore, if we are able to compute a metric distance among the input patterns, we can obtain the corresponding similarity/kernel matrix, $\mathbf{K}$, regardless the nature of the input domain (the converse is also obviously true).
Since in this paper we deal with sequences of variable length, we compute the input data distance by using the Levenshtein distance.
Depending on the specific sequence type, we use static (i.e., predefined) and ad hoc substitution weights for characterizing the edit operation costs. A kernel function is obtained from the Levenshtein string metric according to the aforementioned mathematical expressions.
Correction techniques \cite{Duin2012826} are not used here to make sure that the obtained kernel is positive definite regardless the definition of $d(\cdot, \cdot)$.

Dataset DS-1811, which contains sequences of characters, has been processed with a \textit{kernelized} C-SVM based classifier \cite{schoelkopf+smola2002}. We use three different weighting schemes to process DS-1811.
The first one uses a static cost scheme: all substitutions have unitary costs.
In the second case we use the substitution costs given by the PAM120 matrix \cite{sub_matrix}.
In the latter setting, we determine the substitution costs as the solution of a problem-dependent, global optimization described in the following.
Let $\mathbf{S}\in[0, 1]^{20\times 20}$ be a non-negative, zero-diagonal, cost matrix which contains at position $S_{ij}$ the cost of substituting the \textit{i}-th amino acid with the \textit{j}-th.
The best-performing $\mathbf{S}$ is determined as the cost matrix that maximizes the classification accuracy achieved on a suitable control dataset.
The global optimization is driven by the performance of the classifier on the dataset instance, and it is implemented here by means of a standard genetic algorithm.
In practice, we search for a custom solution, i.e., a matrix $\mathbf{S}$ that is specifically tuned for the particular problem at hand.
Each substitution matrix is formally encoded as a real-valued vector lying in $[0, 1]^{400}$.
Selection, cross-over, and random mutation operators of the genetic algorithm are applied on such vectors to let the population of candidate solutions (i.e., cost matrix instances) evolve over the iterations.
The stop criterion used to determine the convergence of the genetic algorithm is implemented by checking for a maximum number of iterations (100) and by evaluating if the fitness has not changed during the last 10 iterations.

In DS-G-454, instead, we process input patterns that are labeled graphs. Therefore, we implement the kernel function as the graph coverage kernel \cite{livi2012gc,livi2012_pgm}.
DS-454 is a subset of DS-1811, and therefore the same classification systems described for DS-1811 apply here.
Finally, DS-S-454 is again a dataset of sequences, although the objects in the sequences are three-dimensional real-valued vectors; therefore in this case we use the Euclidean distances among such vectors for the substitution weights of the Levenshtein distance.

\subsection{Global structural complexity measures for the protein graphs}
\label{sec:complexity}

Besides their huge differences in shape, proteins contact maps (graphs) share a common modular pattern \cite{tasdighian2013modules,mixbionets2}. The application of Guimer\`{a} and Amaral network cartography \cite{guimera2005cartography} on a large set of proteins, demonstrated the existence of a common role subdivision among amino acid residues for all protein molecules \cite{tasdighian2013modules}. This suggests a common intrinsic complexity of protein graphs wiring pattern.
The basic invariance of protein graph complexity re-assures us of the possibility of relying on a common scaling and thus on the homogeneity of the considered objects, upon which pattern recognition is applied.

To provide experimental evidence of the conjecture that folded proteins ``look all the same'' from the pure topological viewpoint \cite{doi:10.1021/cr3002356}, we analyzed the complexity of the graphs in DS-G-454. The complexity of a graph topology, which is a well-known and widely discussed concept in the scientific literature \cite{Dehmer201157,havlin2010}, have been measured for the data in DS-G-454 in two complementary ways: by (i) calculating the 2-order R\'{e}nyi entropy \cite{rrnyi1961measures} of the associated Markov chain stationary distribution \cite{norris98Markovchains}, and (ii) computing the recently-proposed graph ambiguity measure \cite{Livi_ga_2013}, which is rooted in the fuzzy sets context.
Since the two methods are conceived in two different mathematical settings, we effectively analyze the graph complexity in terms of two different interpretations of the same concept: the uncertainty. In fact, the first method provides an interpretation of the uncertainty of a graph in terms of randomness (transition probability among the states--vertices), while the second method gives us an evaluation of the ambiguity, which refers instead to the degree of irregularity of the graph's topology.

\subsubsection{Uncertainty of a random walk in the graph}
\label{sec:2order_entropy}

Let $G=(\mathcal{V}, \mathcal{E}), |\mathcal{V}|=n$, be an undirected graph. The Markov chain associated to $G$ is completely described by the graph transition matrix, $\mathbf{T}^{n\times n}$.
The stationary probability distribution (also called equilibrium distribution) of the chain is a probability vector $\pi\in[0, 1]^n, \sum_{i=1}^{n} \pi_i = 1$, such that $\pi\mathbf{T}=\pi$; it can be seen also as the left eigenvector of $\mathbf{T}$ with eigenvalue 1.
The stationary distribution $\pi$ always exists but in general it is not unique; moreover, it depends on the initial distribution and on other properties of the chain \cite{Lovasz1996}. A straightforward and always valid way to calculate a stationary distribution of a graph $G$ is to set the initial distribution of the chain as $\pi$, defined as follows:
\begin{equation}
\pi_i = \frac{\left( \sum_{j=1}^{n} A_{ij} \right)}{2|\mathcal{E}|}.
\end{equation}

Given $\pi$, we computed the $\alpha$-order R\'{e}nyi entropy, with $\alpha=2$, to evaluate the complexity.
The 2-order R\'{e}nyi entropy is computed as:
\begin{equation}
H(\pi) = -\log\left(\sum_{i=1}^{n} \pi_{i}^{2}\right) .
\end{equation}

It is well-known that $H(\pi)$ is upper bounded by $\log(n)$ and therefore it can be normalized in the unit interval. The quantity approaches one as the distribution becomes uniform (i.e., maximally uncertain--complex), while it tends to zero as it converges to the degenerate distribution. A $d$-regular graph, for instance, is associated with a uniform stationary distribution of the states (all states--vertices are equally likely to be visited).

\subsubsection{Ambiguity of the graphs}
\label{sec:ambiguity}

The ambiguity of a graph $G=(\mathcal{V}, \mathcal{E}), |\mathcal{V}|=n$, gives a measure of uncertainty elaborated according to a fuzzy set based interpretation \cite{Livi_ga_2013}.
The ambiguity of the graph $G$ is calculated by embedding the graph into a fuzzy hypercube $\mathcal{I}=[0, 1]^n$, which, in short, encodes the membership values of the vertices.
A graph $G$ is mapped to a type-1 fuzzy set $\mathcal{F}$, defined as
\begin{equation}
\mathcal{F}=\{(v, \mu_{\mathcal{F}}(v)) | \ v\in\mathcal{V}(G), \mu_{\mathcal{F}}(v)\in[0, 1] \},
\end{equation}
by generating the membership function $\mu_{\mathcal{F}}(\cdot)$ of the graph vertices, $\mathcal{V}(G)$.
Such a membership function is constructed by considering the so-called partition $P$ of the graph, which basically is a representation of the vertex set $\mathcal{V}(G)$ as the union of $k$ disjoint subsets $\mathcal{C}_i$ of vertices:
\begin{equation}
P=\bigcup_{i=1}^{k} \mathcal{C}_i.
\end{equation}

The partition $P$ is then fuzzified by computing the t-conorm $\bot$ \cite{pedrycz1998introduction} among the fuzzy sets $\mathcal{F}_i$ associated to each $\mathcal{C}_i$, yielding the resulting fuzzy set $\mathcal{F}$:
\begin{equation}
\label{eq:fuzz}
\mathcal{F}= \displaystyle\bot_{i=1}^{k} \mathcal{F}_i.
\end{equation}

Eq. \ref{eq:fuzz} is synthetically denoted as $\mathcal{F}=\phi^{P}(G)$ in the following.
The membership function $\mu_{\mathcal{F}_{i}}(\cdot)$ describing the fuzzy set $\mathcal{F}_i$ is generated according to the following expression:
\begin{equation}
\mu_{\mathcal{F}_{i}}(v) = \alpha_{\mathcal{C}_{i}}(v) \times \beta_{\mathcal{C}_{i}}(v).
\end{equation}

Notably, $\alpha_{\mathcal{C}_{i}}(v)$ accounts for the degree concentration of vertex $v$ in $\mathcal{C}_i$, while $\beta_{\mathcal{C}_{i}}(v)$ gives the importance of the vertex in $\mathcal{C}_i$ in terms of centrality.
Given the fuzzy set $\mathcal{F}$ representing the uncertainty of the graph $G$, the measure of ambiguity of $G$, denoted as $A(G)$, is obtained by computing (any monotonic non-decreasing transformation of) the fuzzy entropy of $\mathcal{F}$ \cite{pedrycz1998introduction}.
Since there are exponentially-many fuzzy set representations for a single graph $G$, the ambiguity value is calculated as the solution of the following combinatorial optimization problem:
\begin{equation}
A(G) = \min_{P} A(\phi^{P}(G)),
\end{equation}
where $A(G)$ assumes values within the $[0, 1]$ range, approaching one as the graph is maximally ambiguous (i.e., maximally irregular). It has been proved that $A(G)$ is zero when the graph is regular (e.g., complete) \cite{Livi_ga_2013}. Accordingly, $A(G)$ can be used as a global complexity descriptor for the topology of $G$, which in particular characterizes the irregularity of the topology of $G$.

\section{Experimental results and discussion}
\label{sec:results}

In this section, we present the experimental results achieved in this paper. In Section \ref{sec:cca} we first demonstrate the soundness, in terms of statical analysis, of the data used for the classification of the E. coli proteome. In Section \ref{sec:class_results} we show the test set classification accuracy results obtained on the alternative protein representations herein considered.
Finally, in Section \ref{sec:complexity_results} we analyze the complexity of the graphs representing the (available) E. coli proteins.

\subsection{Canonical correlation analysis}
\label{sec:cca}

In order to assess the chemico-physical relevance of the obtained classification performance, we checked the possibility to project the SVM solution on DS-1811 by means of the learned cost matrix $\mathbf{S}$ -- that was based only on literal information of residues -- into a chemico-physical consistent solution.
To reach this goal, a combined principal component/canonical analysis strategy was adopted \cite{cca}, which is described in the following.

The available 243 chemico-physical descriptors \cite{apdbase} were submitted to a principal component analysis (PCA), ending up in a three-component solution, globally explaining 70.2\% of total variation (PC1 = 42.3\%, PC2 = 16.3\%, and PC3 = 11.6\%).
The three extracted components allowed for a straightforward interpretation: PC1 is a hydrophobicity score (positive correlation with hydrophobicity scales, negative correlation with hydrophilicity scales); PC2 is a size/steric hindrance component (very high loadings with volume / size scales); finally, PC3 can be considered as a ``rigidity'' scale (high positive loading with turn propensity, negative loadings with alpha-helix propensity scales).
Those three components allow us to generate a matrix having as rows the 20 amino acid species and as columns the three correspondent ChemPhys1 (Hydrophobicity), ChemPhys2 (Size), and ChemPhys3 (Structural Rigidity) component scores.

An analogous approach was followed on the calculated best-performing cost matrix, $\mathbf{S}$, corresponding to the SVM-based solution obtained by considering the classification performance as feedback. 
We computed the three principal components of $\mathbf{S}$, having as rows the different amino acids and as columns the ``cost'' of their transformation into different amino acid species.
In this case, a seven-components solution was obtained, explaining 68.8\% of total variability. In the following, we refer to those components with SVM1--SVM7.

We are now in the condition to look for the existence of a sound correlation structure linking the chemico-physical and substitution cost spaces.
Mathematically, this problem translates into a canonical correlation analysis. The linear combinations of the chemico-physical components and the cost matrix components are then computed, with the objective of maximizing the linear correlation. These best correlated combinations are called canonical variates.
They are extracted by the algorithm in order of correlation magnitude; moreover, they are mutually orthogonal to each other, so describing independent ``transformation rules'' between the two heterogeneous spaces.
The fact that we use as original variables principal component scores, which are each other mutually independent inside their respective spaces, reassures us that any observed correlation will be driven by inter-fields correlations only.
The results of the canonical correlation analysis are reported in Table \ref{tab:table1}.
\begin{table}[tph!]\footnotesize
\caption{Results of canonical correlation analysis.}
\begin{center}
\begin{tabular}{|c|c|c|c|c|}
\hline
\rowcolor{lgray} & \textbf{Canonical Correlation (CC)} & \textbf{Adjusted CC} & \textbf{Approx Std Err} & \textbf{Squared CC} \\
\hline
\cellcolor{lgray}\textbf{1} & 0.901527 & 0.856986 & 0.042958 & 0.812751 \\
\hline
\cellcolor{lgray}\textbf{2} & 0.747438 & 0.666824 & 0.101250 & 0.558663 \\
\hline
\cellcolor{lgray}\textbf{3} & 0.468202 & 0.333368 & 0.179125 & 0.219213 \\
\hline
\end{tabular}
\label{tab:table1}
\end{center}
\end{table}
It is possible to note two statistically significant canonical variates: the first one scoring a squared canonical correlation around 0.8 (with $r=0.9$), while the second canonical variate gives $r=0.75$, correspondent to a model accounting for 55\% of explained variance (R-square).
In Table \ref{tab:table2} and Table \ref{tab:table3}, respectively, are reported the correlation among chemico-physical and substitution cost variables with the respective canonical variates; the significant correlations are bolded.
\begin{table}[tph!]
\begin{center}
\caption{Correlations among the chemico-physical variables and their canonical variates.}
\begin{tabular}{|c|c|c|c|}
\hline
\rowcolor{lgray} & \textbf{Variate 1} & \textbf{Variate 2} & \textbf{Variate 3} \\
\hline
\cellcolor{lgray}\textbf{ChemPhys1} & -0.1632 & \textbf{0.9759} & -0.1447 \\
\hline
\cellcolor{lgray}\textbf{ChemPhys2} & \textbf{0.8269} & 0.2153 & 0.5194 \\
\hline
\cellcolor{lgray}\textbf{ChemPhys3} & \textbf{-0.5381} & 0.0349 & 0.8422 \\
\hline
\end{tabular}
\label{tab:table2}
\end{center}
\end{table}
\begin{table}[tph!]
\begin{center}
\caption{Correlations between the SVM variables and their canonical variates.}
\begin{tabular}{|c|c|c|c|}
\hline
\rowcolor{lgray} & \textbf{Variate 1} & \textbf{Variate 2} & \textbf{Variate 3} \\
\hline
\cellcolor{lgray}\textbf{SVM1} & \textbf{0.4465} & -0.0808 & 0.4137 \\
\hline
\cellcolor{lgray}\textbf{SVM2} & 0.3212 & 0.3470 & -0.5716 \\
\hline
\cellcolor{lgray}\textbf{SVM3} & -0.1259 & -0.1012 & 0.3203 \\
\hline
\cellcolor{lgray}\textbf{SVM4} & -0.1851 & -0.0256 & 0.4989 \\
\hline
\cellcolor{lgray}\textbf{SVM5} & \textbf{0.5303} & -0.0620 & 0.2210 \\
\hline
\cellcolor{lgray}\textbf{SVM6} & \textbf{-0.6024} & 0.1849 & 0.0061 \\
\hline
\cellcolor{lgray}\textbf{SVM7} & -0.0566 & \textbf{-0.9078} & -0.3189 \\
\hline
\end{tabular}
\label{tab:table3}
\end{center}
\end{table}

The most relevant (first canonical variate) transfer rule linking chemico-physical and discrimination spaces is mainly driven by amino acid size (ChemPhys1) acting in opposition with ``rigidity'' component (ChemPhys3).
The second canonical variate, instead, is linked to the unique role of hydrophobicity, but this is a minor effect (only the seventh component of the discrimination space scales significantly with the canonical variate).
Cost matrix reports the ``expense'' of a given amino acid substitution: high costs (near 1) imply a significant change in the classification, while low costs (near zero) the relative indifference of the substitution in terms of effects on classification.
The presence of significant canonical correlations between chemico-physical features and amino acids substitution costs implies that the hidden discrimination logic is rooted on justifiable chemico-physical properties.

However, by no means we can derive ``general linear models'' based on single amino acid features, because the same move (e.g., substituting residue A with V) has an opposite effect in terms of ``soluble/not soluble'', depending on the starting point.
Moreover, the different moves have both positive and negative loadings on different SVM components. This means that we cannot derive any global aggregation or folding value for a given move: the effect of each move depends on the starting point and has a different value on different eigenvectors, thus a move can be at a high cost for SVM3 and low cost for SVM4.
The SVM components, corresponding to the hidden rules at the basis of cost matrix $\mathbf{S}$, are organized in terms of variance explained.
This implies that SVM1 is more relevant than SVM2. The resulting cost is a sort of majority vote over the components, and it can by no means be collapsed to simple general rules expressed in context independent chemico-physical terms without further specification.
This explains both the lack of any relevant general correlation between average physical properties and solubility classification, and the extreme sensitivity of solubility/aggregation equilibrium by single point mutations.

\begin{figure}[ht!]
 \centering
 \includegraphics[viewport=0 0 476 346,scale=0.75,keepaspectratio=true]{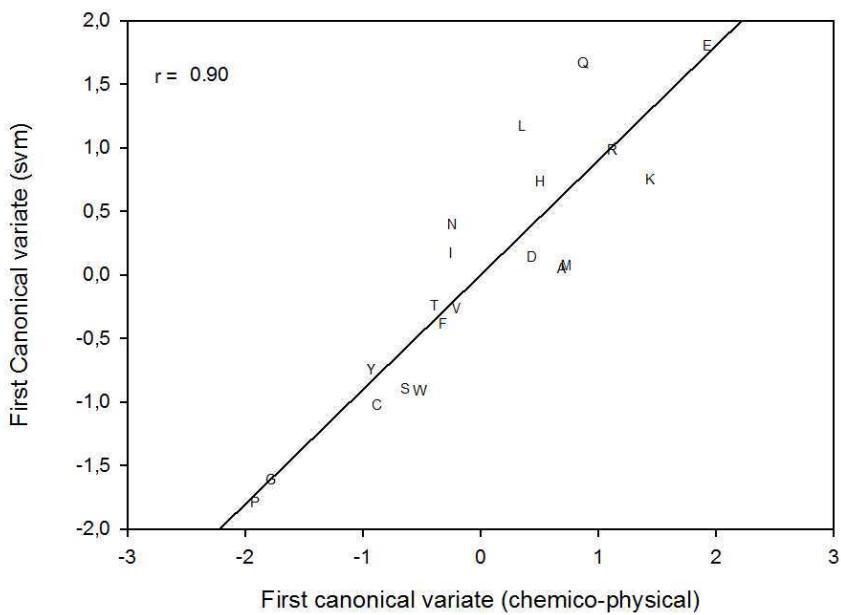}
 \caption{Correlation analysis among the two most important canonical variates.}
 \label{fig:correlation}
\end{figure}

In Figure \ref{fig:correlation}, the plot showing the relation between chemico-physical and SVM spaces in terms of canonical variates is reported, keeping in mind that SVM variates refer to a ``derivative'' space.
That is, the entity of the effect exerted by a single amino acid changes on its solubility and not the general soluble/insoluble propensity of amino acid residues that, as we explained, does not exist as such. The plot can thus be intended as a proof of the chemico-physical relevance of the solution found by the proposed algorithm -- i.e., the best-performing matrix $\mathbf{S}$ -- and thus of its scientific soundness.

\subsection{Classification accuracy results}
\label{sec:class_results}

In their paper, \citet{niwa2009} noted that soluble proteins were considerably smaller than insoluble ones.
This is in line with many results coming from other groups \cite{Waldo200333,PRO:PRO122057,Kubelka200476} and with the well-known dependence of folding process from molecules size (\citet{PRO:PRO122057}).
In order to have a baseline with which contrasting our results in terms of recognition, we developed a discriminator operating on the basis of protein size only, i.e., the number of residues in the protein.
Tests have been performed on DS-1811. The best discriminating threshold was found at 246 residues: a protein containing less than 246 residues is deterministically classified as soluble.
Such a classification scheme commits 301 errors for the insoluble proteins (error rate of 0.198) and 33 for the soluble (error rate of 0.3).
This implies a marked tendency of small proteins to be more soluble than large ones \cite{ramshini2011large}.
The fact that the best discrimination threshold corresponds to 246 residues length is consistent with the results outlined by \citet{PROT:PROT21179} that, on the sole basis of between-residues contact density as a function of protein length, discovered a maximum preferred size for protein domains located between 180 and 320 residues length with a peak at 274.
In the analysis reported in \citet{PROT:PROT21179}, no solubility data were specifically addressed; they were solely based on the number of protein internal contacts. Moreover, the results were demonstrated to be very general and independent of the analyzed dataset.
The fact that the present analysis, in a completely independent manner, re-discovers the same threshold is of extreme importance, especially if we consider that the density of within-molecule contacts exponentially decreases at increasing length until approximately 250-300 residues length, and that from this threshold onward the density of contacts remains approximately constant.
The higher density of within-molecule contacts corresponds to a preference for within-molecule interactions (and consequently toward the soluble pole). This gives a simple, albeit very raw, rationalization to the observed convergence toward the same threshold of contact density scaling and soluble/insoluble threshold.
On a more general perspective, this result tells us that the consideration of topological features (contact graphs) of protein 3D structure can be a useful perspective to be adopted for the discrimination of soluble/insoluble proteins.

Let us now move to the results obtained by means of the various SVM-based classifiers.
The results reported in Table \ref{tab:testset_results_ecoli} have been obtained by setting $C=2$ for C-SVM; such a setting has been obtained by preliminary tests.
Overall, the comparison with respect to the aforementioned baseline discrimination approach is in favour of C-SVM.
Results on DS-1811 (sequence-based analysis) show satisfactory error rates for the two classes of insoluble and soluble proteins, regardless the weighting scheme adopted for the Levenshtein distance. Note that the row of this table indicating ``EVOLUTIVE'' as the substitution matrix is the solution that adopts the aforementioned best-performing substitution cost matrix, $\mathbf{S}$.

Results on DS-G-454, i.e., the dataset of labeled graphs, show considerably worse results. In fact, the error rate for the soluble class approaches 0.8, denoting overall an unreliable classification.
Error rates on DS-454 denote a considerable drop of performances for the recognition of the soluble class with respect to DS-1811. Since the dataset of DS-454 is composed only of those sequences for which we have the corresponding graph representations in DS-G-454, we deduce that such a subset of proteins induces a more complex problem than the one on DS-1811 from the pure class discrimination viewpoint; this is also substantiated by the reduced number of available patterns.
On the other hand, results for DS-S-454 are interesting, considering both per-class and global error rates.
We stress that this dataset has been obtained by the seriation of the graph representations in DS-G-454, which is performed by using the information of the first eigenvector of the transition matrix.
Therefore, such an approach could be considered as a graph-based pattern recognition technique, even if the classification is performed by dealing with sequences of vectors.
To further strengthen the results obtained on DS-S-454, we repeated the experiment by randomizing the class labels of the training set patterns; test set is left unaltered. As expected, test set results are much worse than the ones presented in Tab. \ref{tab:testset_results_ecoli}.
In fact, the global error rate drops to 0.5471, with 75 and 12 errors for the (randomized) insoluble and soluble classes, respectively.

The important differences of results achieved on DS-G-454 and DS-S-454, by considering the adopted computational systems, are not easy to justify from a pure algorithmic viewpoint.
That is, the precise reason why operating directly on the labeled graph space yields worse results are not easy to rationalize, since both approaches (i.e., DS-G-454 and DS-S-454) operate basically on the same structural and chemico-physical information, although arranged in two different settings (i.e., respectively graph and sequence).
However, as recently pointed out \cite{PROT:PROT22113}, a subset of the eigenvectors of the contact (adjacency) matrix provides sound descriptors of the protein structure, showing good correlation with hydrophobicity.
It is worth to stress that, since the classification approach used for DS-S-454 effectively makes explicit use of the eigenvalues/eigenvectors of the transition matrix (which is an elaboration of the contact/adjacency matrix) to define the order of the graph vertices in the sequence, such a technique is well-justified from the biological viewpoint, taking also into account chemico-physical information derived by the previously discussed statistical analysis.
\begin{table}[tph!]\footnotesize
\begin{center}
\caption{Test set classification accuracy results achieved on the considered datasets.}
\begin{tabular}{|c|p{1.95cm}|p{2.4cm}|p{2.3cm}|p{2.5cm}|p{1.7cm}|}
\hline
\rowcolor{lgray}\textbf{DS} & \textbf{Subst. Matrix} & \textbf{\# ERR. INS. / \# INS.} & \textbf{\# ERR. SOL. / \# SOL.} & \textbf{ERR. RATE INS. -- ERR. RATE SOL.} & \textbf{GLOBAL ERR. RATE} \\
\hline
\multirow{3}{*}{DS-1811}
& -- & 232/1521 & 33/110 & 0.1525 -- 0.3000 & 0.1624 \\
\cline{2-6}
& PAM120 & 246/1521 & 28/110 & 0.1617 -- 0.2545 & 0.1679 \\
\cline{2-6}
& EVOLUTIVE & 304/1521 & 22/110 & 0.1998 -- 0.2000 & 0.1998 \\
\hline
DS-G-454 & -- & 12/132 & 21/27 & 0.0909 -- 0.7777 & 0.2075 \\
\hline
DS-454 & -- & 4/132 & 20/27 & 0.0303 -- 0.7407 & 0.1509 \\
\hline
DS-S-454 & -- & 10/132 & 11/27 & 0.0757 -- 0.4074 & 0.1320 \\
\hline
\end{tabular}
\label{tab:testset_results_ecoli}
\end{center}
\end{table}

\subsection{Evaluation of protein graph structural complexity}
\label{sec:complexity_results}

We calculated the normalized 2-order R\'{e}nyi entropy (as described in Section \ref{sec:2order_entropy}) for every graph in DS-G-454. As expected, the graphs have almost identical entropy (i.e., complexity), regardless the soluble/insoluble original classification. The insoluble proteins have average entropy of 0.9784, with standard deviation 0.0057; the soluble ones have average entropy 0.9726 and standard deviation 0.0084.
As for the 2-order R\'{e}nyi entropy, we calculated also the graph ambiguity (Section \ref{sec:ambiguity}) of the same graphs in DS-G-454. Analogously, the ambiguity does not seem to provide an effective discriminating feature among soluble and insoluble proteins (although it gives a slightly better average discrimination than the 2-order R\'{e}nyi entropy).
In fact, the insoluble proteins have average ambiguity of 0.1745, with standard deviation 0.0221; the soluble ones have average ambiguity of 0.1980 with standard deviation 0.0313.

\begin{figure}[ht!]
 \centering
 \includegraphics[viewport=0 0 349 243,scale=0.8,keepaspectratio=true]{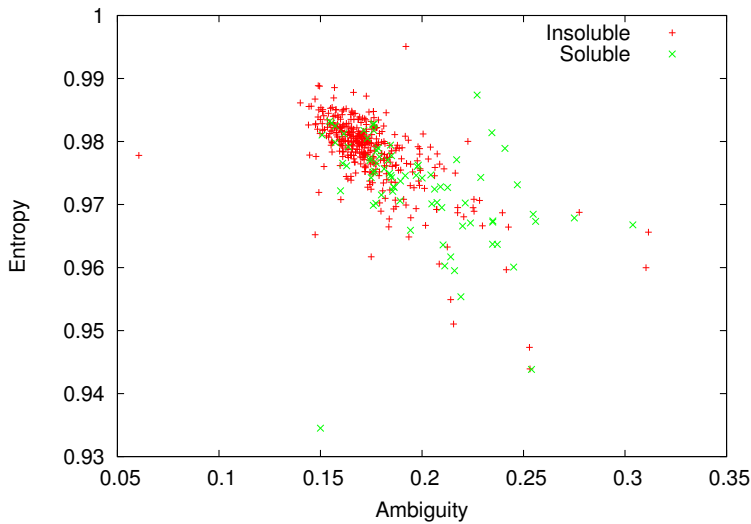}
 \caption{Data distribution of the 454 graphs represented as two-dimensional vectors containing the corresponding calculated entropy and ambiguity values.}
 \label{fig:complex_distr}
\end{figure}

According to herein considered interpretations of the structural complexity of a graph, such results points toward the fact that proteins have a common global topological architecture.
It is worth mentioning that we performed additional experiments by representing each graph in the DS-G-454 dataset as a feature vector consisting of two real-valued features: the calculated values of entropy and ambiguity.
The related data distribution is shown in Figure \ref{fig:complex_distr}. We considered a C-SVM classifier operating with a ``regular'' Gaussian kernel on such two-dimensional vectors, with the aim of formally quantifying the class discrimination performance.
The obtained results (full details are not shown here) demonstrate the inability of discriminating soluble and insoluble proteins when represented according to only those two structural complexity descriptors, achieving an error rate of 100\% for the soluble class on the test set (training and test sets are constructed by considering several instances with the same split percentages of DS-G-454).

Nonetheless, such results cannot be considered as definitive. In future studies we will evaluate additional structural descriptors, enriching this type of vector-based representation of protein graphs with further, more complex characteristics.

\section{Conclusions and future directions}
\label{sec:conclusions}

The aim of this study was the derivation of useful insights on the nature of protein folding/aggregation by following a data-driven paradigm.
We exploited different representations of proteins, all based on a consistent and pure experimental data base \cite{niwa2009}.
The use of sequence and graph based representations of patterns implied the use of pattern recognition methods capable of dealing with non-geometric data.
The problem of aggregation propensity recognition was demonstrated to be solvable at the single molecule scale starting from pure sequence data.
Moreover, the scoring of a statistically significant canonical correlation among computed substitution cost matrix and classical chemico-physical properties of amino acids, assures us of the existence of a still controversial and largely context-dependent ``physical code'' of protein folding \cite{dill2012protein}, which we can interpret as a sort of ``transfer equation'' expressed in chemico-physical terms linking sequence and aggregation propensity spaces.
This transfer equation was demonstrated to arise from the combinations of different mutually orthogonal components rooted on chemico-physical properties of amino acid residues. This is strictly dependent on contextual information consistently with the presence of a very delicate balance between folding and aggregation modes, which could be dramatically shifted even by single mutation events.
Additionally, we also reconfirmed the results in Ref. \cite{Agostini2012237} by using a more parsimonious model, that is, with three chemico-physical descriptive variables instead of 28.

The interplay of different forces in folding (van der Waals force here represented by amino acids size component, hydrophobic interaction, and backbone angle preference here represented by structural rigidity component), confirms the general picture given by \citet{dill2012protein}.
The dramatic effect of protein size upon folding/aggregation balance was demonstrated to be mediated by contact density consistently with the recognized importance of contact order in folding rate (\citet{PRO:PRO122057}) and, more in general, by the relative prevalence of ``within-molecule'' vs. ``between-molecules'' contacts.
It is worth noting that the 246 residues length we demonstrated to be the best threshold length for the soluble/insoluble, baseline discrimination fits very well with the limit where an obliged multi-domain organization starts, as described in \citet{PROT:PROT21179}.
This result constitutes a link between size and topological considerations.

The striking differences among the test set classification accuracy results achieved on DS-S-454, DS-454, and DS-G-454 deserve a further analysis. The DS-454 dataset, which is constituted by the proteins for which the 3D structural information is available as PDB, constitutes a not-so-good sample, as evident by the high rate of error for the soluble class (0.74 vs. 0.20 achieved on DS-1811).
On the base of the results achieved with the adopted technique operating directly on the labeled graphs space, we can conclude that this specific computational approach was not sufficiently effective.
In fact, the system was tested for its discrimination power on the DS-G-454 dataset, obtaining classification accuracy results that are slightly worse that the ones on DS-454; the recognition rate of the soluble class is weak, with an error rate equal to 0.77.
However, such a conclusion regarding the results on DS-G-454 cannot be considered as definitive. In fact, we adopted a SVM-based classification system operating in the labeled graph space by means of a graph kernel function.
However, in the graph-based pattern recognition context there is a plethora of techniques that could be used for such a purpose \cite{gm_survey,odse,Hancock2012833,gralg_2012}.
On the other hand, when the same graph topological information was expressed in another form by ordering the amino acid residues of each protein according to the spectral decomposition of their transition matrix, the discrimination power reached a global error rate of 0.13 (the best general performance achieved in this paper). Moreover, the error rate for the soluble class went down to 0.40, definitely better than the 0.74 and 0.77 achieved on DS-454 and DS-G-454, respectively.
This result tells us that we designed a very interesting type of protein representation (DS-S-454), which allowed us to achieve a reasonable classification accuracy, much higher than the one based on sequence of amino acids identifiers only (i.e., DS-454).

\citet{PROT:PROT22113} predicted such efficiency of multilevel representation of protein topology obtained by the projection of topological features on protein sequence.
The Effective Connectivity (EC) profile \cite{PROT:PROT20240} the authors proposed is related to our approach, being based on the eigenvalues of the proteins contact maps.
Bastolla et al. explicitly state ``...the EC profile allows to define a natural measure of modularity that correlates with the number of domains composing the protein...'' and ``...structurally similar proteins have very similar EC profiles, so that the similarity between aligned EC profiles can be used as a structural similarity measure...''.
In their work, the authors complement the pure topological EC with hydrophobicity information showing a marginal but statistically significant improvement in the ``reverse folding problem'' of predicting sequence order by the EC profile.

Here we showed compatible results with the Bastolla et al. proposal, inserting a more refined chemico-physical description of residues (the first three chemico-physical components obtained by the PCA) and imposing an evolution-motivated substitution metrics, obtaining a very good discrimination power as for the hard task of aggregation/folding discrimination.
At odds with Bastolla et al., we adopted a transition matrix based graph representation instead of contact (i.e., adjacency) matrix.
This choice was motivated by the goal of obtaining a global structural complexity measure characterizing the protein 3D structures (based on the stationary distribution of the associated Markov chain). This fact was based on the hypothesis that the more complex the protein architecture is, the higher the difficulty of reaching a stable folding state.
To this end, we also reported preliminary experimental results that showed the high similarity of protein topology, regardless their solubility degree and number of residues.
Nonetheless, more effort will be devoted in future studies to this specific aspect of analyzing the protein graph complexity, for the purpose of discrimination and prediction of the solubility degree.

There are plenty of results (for a review see \citet{doi:10.1021/cr3002356}) pointing to the fact that protein contact maps play, in the realm of proteins, the same role played by structural formula in organic chemistry.
The apparently dramatic loss of information consequent to shift from the whole-rank structural information (the 3D coordinates of each single atom) to the yes/no information of effective contacts between alpha carbons, ends up in the focusing on the actually relevant part of information analogously to what happens with chemical graphs of small organic molecules, where the simple consideration of structural formula was demonstrated to be much more efficient than more sophisticated approaches for predicting their biological activity \cite{bender2005discussion}.
Here in this paper we gave another proof of the central role played in protein science by the use of topological information combined with chemico-physical descriptions of residues.
In our opinion, these results points toward the realization of a consistent and multilevel formalization of protein molecules integrating heterogeneous information.

On a more general ground, we demonstrated the possibility to derive valuable information on the nature of a given investigated phenomenon, operating on the different discrimination ability of pattern recognition techniques as applied to alternative representations of the same objects.
This fact opens the way to the consideration of pattern recognition methods as measurement devices, not so different from a microscope or a spectrophotometer, so closing the gap between computational and content-oriented classical scientific approaches.
In addition, the combination of the the information coming from both wiring architecture and features of specific vertices into a synthetic descriptive frame can be important as a general quantification scheme for any classification problem involving networked systems.

\bibliographystyle{abbrvnat}
\bibliography{/home/lorenzo/University/Research/Publications/Bibliography.bib}

\begin{thebibliography}{46}
\providecommand{\natexlab}[1]{#1}
\providecommand{\url}[1]{\texttt{#1}}
\expandafter\ifx\csname urlstyle\endcsname\relax
  \providecommand{\doi}[1]{doi: #1}\else
  \providecommand{\doi}{doi: \begingroup \urlstyle{rm}\Url}\fi

\bibitem[apd()]{apdbase}
{Amino acid Physical-chemical property Database}.
\newblock URL \url{http://www.rfdn.org/bioinfo/APDbase/index.html}.

\bibitem[pdb()]{pdb}
{Protein Data Bank}.
\newblock URL \url{http://www.rcsb.org/pdb/home/home.do}.

\bibitem[sub()]{sub_matrix}
{Substitution Matrix}.
\newblock URL \url{ftp://ftp.ncbi.nih.gov/blast/matrices}.

\bibitem[Agostini et~al.(2012)Agostini, Vendruscolo, and
  Tartaglia]{Agostini2012237}
F.~Agostini, M.~Vendruscolo, and G.~G. Tartaglia.
\newblock {Sequence-Based Prediction of Protein Solubility}.
\newblock \emph{Journal of Molecular Biology}, 421\penalty0 (2-3):\penalty0
  237--241, 2012.
\newblock ISSN 0022-2836.
\newblock \doi{10.1016/j.jmb.2011.12.005}.

\bibitem[Bastolla et~al.(2005)Bastolla, Porto, Roman, and
  Vendruscolo]{PROT:PROT20240}
U.~Bastolla, M.~Porto, H.~E. Roman, and M.~Vendruscolo.
\newblock {Principal eigenvector of contact matrices and hydrophobicity
  profiles in proteins}.
\newblock \emph{Proteins: Structure, Function, and Bioinformatics}, 58\penalty0
  (1):\penalty0 22--30, 2005.
\newblock ISSN 1097-0134.
\newblock \doi{10.1002/prot.20240}.

\bibitem[Bastolla et~al.(2008)Bastolla, Ort{\'i}z, Porto, and
  Teichert]{PROT:PROT22113}
U.~Bastolla, A.~R. Ort{\'i}z, M.~Porto, and F.~Teichert.
\newblock {Effective connectivity profile: A structural representation that
  evidences the relationship between protein structures and sequences}.
\newblock \emph{Proteins: Structure, Function, and Bioinformatics}, 73\penalty0
  (4):\penalty0 872--888, 2008.
\newblock ISSN 1097-0134.
\newblock \doi{10.1002/prot.22113}.

\bibitem[Bender and Glen(2005)]{bender2005discussion}
A.~Bender and R.~C. Glen.
\newblock {A discussion of measures of enrichment in virtual screening:
  comparing the information content of descriptors with increasing levels of
  sophistication}.
\newblock \emph{Journal of Chemical Information and Modeling}, 45\penalty0
  (5):\penalty0 1369--1375, 2005.
\newblock \doi{10.1021/ci0500177}.

\bibitem[Bianchi et~al.(2014)Bianchi, Livi, Rizzi, and Sadeghian]{gralg_2012}
F.~M. Bianchi, L.~Livi, A.~Rizzi, and A.~Sadeghian.
\newblock A {G}ranular {C}omputing approach to the design of optimized graph
  classification systems.
\newblock \emph{Soft Computing}, 18\penalty0 (2):\penalty0 393--412, 2014.
\newblock ISSN 1432-7643.
\newblock \doi{10.1007/s00500-013-1065-z}.

\bibitem[Chiti and Dobson(2006)]{doi:10.1146/annurev.biochem.75.101304.123901}
F.~Chiti and C.~M. Dobson.
\newblock {Protein Misfolding, Functional Amyloid, and Human Disease}.
\newblock \emph{Annual Review of Biochemistry}, 75\penalty0 (1):\penalty0
  333--366, 2006.
\newblock \doi{10.1146/annurev.biochem.75.101304.123901}.
\newblock PMID: 16756495.

\bibitem[Chong et~al.(2011)Chong, Lee, Kang, Park, and
  Ham]{doi:10.1021/ja1116233}
S.-H. Chong, C.~Lee, G.~Kang, M.~Park, and S.~Ham.
\newblock {Structural and Thermodynamic Investigations on the Aggregation and
  Folding of Acylphosphatase by Molecular Dynamics Simulations and Solvation
  Free Energy Analysis}.
\newblock \emph{Journal of the American Chemical Society}, 133\penalty0
  (18):\penalty0 7075--7083, 2011.
\newblock \doi{10.1021/ja1116233}.

\bibitem[Dehmer and Mowshowitz(2011)]{Dehmer201157}
M.~Dehmer and A.~Mowshowitz.
\newblock {A history of graph entropy measures}.
\newblock \emph{Information Sciences}, 181\penalty0 (1):\penalty0 57--78, 2011.
\newblock ISSN 0020-0255.
\newblock \doi{10.1016/j.ins.2010.08.041}.

\bibitem[Deza and Deza(2009)]{Deza.Deza2009EncyclopediaofDistances}
M.~M. Deza and E.~Deza.
\newblock \emph{{Encyclopedia of Distances}}.
\newblock Springer Berlin Heidelberg, Berlin, Germany, 2009.

\bibitem[{Di Paola} et~al.(2012){Di Paola}, {De Ruvo}, Paci, Santoni, and
  Giuliani]{doi:10.1021/cr3002356}
L.~{Di Paola}, M.~{De Ruvo}, P.~Paci, D.~Santoni, and A.~Giuliani.
\newblock Protein contact networks: an emerging paradigm in chemistry.
\newblock \emph{Chemical Reviews}, 113\penalty0 (3):\penalty0 1598--1613, 2012.
\newblock \doi{10.1021/cr3002356}.

\bibitem[Dill and MacCallum(2012)]{dill2012protein}
K.~A. Dill and J.~L. MacCallum.
\newblock {The protein-folding problem, 50 years on}.
\newblock \emph{Science}, 338\penalty0 (6110):\penalty0 1042--1046, 2012.
\newblock \doi{10.1126/science.1219021}.

\bibitem[Duin and P\c{e}kalska(2012)]{Duin2012826}
R.~P.~W. Duin and E.~P\c{e}kalska.
\newblock {The dissimilarity space: Bridging structural and statistical pattern
  recognition}.
\newblock \emph{Pattern Recognition Letters}, 33\penalty0 (7):\penalty0
  826--832, 2012.
\newblock ISSN 0167-8655.
\newblock \doi{10.1016/j.patrec.2011.04.019}.

\bibitem[Foggia et~al.(2014)Foggia, Percannella, and Vento]{foggia2012graph}
P.~Foggia, G.~Percannella, and M.~Vento.
\newblock Graph matching and learning in pattern recognition in the last 10
  years.
\newblock \emph{International Journal of Pattern Recognition and Artificial
  Intelligence}, 28\penalty0 (1):\penalty0 1450001, 2014.
\newblock \doi{10.1142/S0218001414500013}.

\bibitem[Guimer{\`a} and Amaral(2005)]{guimera2005cartography}
R.~Guimer{\`a} and L.~A.~N. Amaral.
\newblock {Cartography of complex networks: modules and universal roles}.
\newblock \emph{Journal of Statistical Mechanics: Theory and Experiment},
  2005\penalty0 (02):\penalty0 P02001, 2005.

\bibitem[Hancock and Wilson(2012)]{Hancock2012833}
E.~R. Hancock and R.~C. Wilson.
\newblock {Pattern analysis with graphs: Parallel work at Bern and York}.
\newblock \emph{Pattern Recognition Letters}, 33\penalty0 (7):\penalty0
  833--841, 2012.
\newblock ISSN 0167-8655.
\newblock \doi{10.1016/j.patrec.2011.08.012}.

\bibitem[Havlin(2010)]{havlin2010}
S.~Havlin.
\newblock \emph{{Complex Networks}}.
\newblock Cambridge University Press, Cambridge, UK, 2010.
\newblock ISBN 9780521841566.

\bibitem[Ivankov et~al.(2003)Ivankov, Garbuzynskiy, Alm, Plaxco, Baker, and
  Finkelstein]{PRO:PRO122057}
D.~N. Ivankov, S.~O. Garbuzynskiy, E.~Alm, K.~W. Plaxco, D.~Baker, and A.~V.
  Finkelstein.
\newblock {Contact order revisited: Influence of protein size on the folding
  rate}.
\newblock \emph{Protein Science}, 12\penalty0 (9):\penalty0 2057--2062, 2003.
\newblock ISSN 1469-896X.
\newblock \doi{10.1110/ps.0302503}.

\bibitem[Kubelka et~al.(2004)Kubelka, Hofrichter, and Eaton]{Kubelka200476}
J.~Kubelka, J.~Hofrichter, and W.~A. Eaton.
\newblock {The protein folding 'speed limit'}.
\newblock \emph{Current Opinion in Structural Biology}, 14\penalty0
  (1):\penalty0 76--88, 2004.
\newblock ISSN 0959-440X.
\newblock \doi{10.1016/j.sbi.2004.01.013}.

\bibitem[Livi and Rizzi(2012)]{livi2012_pgm}
L.~Livi and A.~Rizzi.
\newblock Parallel algorithms for tensor product-based inexact graph matching.
\newblock In \emph{{Proceedings of the 2012 International Joint Conference on
  Neural Networks}}, pages 2276--2283, June 2012.
\newblock ISBN 978-1-4673-1489-3.
\newblock \doi{10.1109/IJCNN.2012.6252681}.

\bibitem[Livi and Rizzi(2013{\natexlab{a}})]{Livi_ga_2013}
L.~Livi and A.~Rizzi.
\newblock {Graph ambiguity}.
\newblock \emph{Fuzzy Sets and Systems}, 221:\penalty0 24--47,
  2013{\natexlab{a}}.
\newblock ISSN 0165-0114.
\newblock \doi{10.1016/j.fss.2013.01.001}.

\bibitem[Livi and Rizzi(2013{\natexlab{b}})]{gm_survey}
L.~Livi and A.~Rizzi.
\newblock The graph matching problem.
\newblock \emph{Pattern Analysis and Applications}, 16\penalty0 (3):\penalty0
  253--283, 2013{\natexlab{b}}.
\newblock ISSN 1433-7541.
\newblock \doi{10.1007/s10044-012-0284-8}.

\bibitem[Livi et~al.(2012{\natexlab{a}})Livi, {Del Vescovo}, and
  Rizzi]{livi+delvescovo+rizzi_seriation+gradis}
L.~Livi, G.~{Del Vescovo}, and A.~Rizzi.
\newblock Graph recognition by seriation and frequent substructures mining.
\newblock In \emph{{Proceedings of the First International Conference on
  Pattern Recognition Applications and Methods}}, volume~1, pages 186--191,
  Feb. 2012{\natexlab{a}}.
\newblock ISBN 978-989-8425-98-0.
\newblock \doi{10.5220/0003733201860191}.

\bibitem[Livi et~al.(2012{\natexlab{b}})Livi, {Del Vescovo}, and
  Rizzi]{livi2012gc}
L.~Livi, G.~{Del Vescovo}, and A.~Rizzi.
\newblock Inexact graph matching through graph coverage.
\newblock In \emph{{Proceedings of the First International Conference on
  Pattern Recognition Applications and Methods}}, volume~1, pages 269--272,
  Feb. 2012{\natexlab{b}}.
\newblock ISBN 978-989-8425-98-0.
\newblock \doi{10.5220/0003732802690272}.

\bibitem[Livi et~al.(2013{\natexlab{a}})Livi, Bianchi, Rizzi, and
  Sadeghian]{odse2_ijcnn_2013}
L.~Livi, F.~M. Bianchi, A.~Rizzi, and A.~Sadeghian.
\newblock Dissimilarity space embedding of labeled graphs by a clustering-based
  compression procedure.
\newblock In \emph{{Proceedings of the 2013 International Joint Conference on
  Neural Networks}}, pages 1646--1653, Dallas, USA, Aug. 2013{\natexlab{a}}.
\newblock ISBN 978-1-4673-6129-3.
\newblock \doi{10.1109/IJCNN.2013.6706937}.

\bibitem[Livi et~al.(2013{\natexlab{b}})Livi, {Del Vescovo}, and
  Rizzi]{seriation+gradis_lncs_2012}
L.~Livi, G.~{Del Vescovo}, and A.~Rizzi.
\newblock Combining graph seriation and substructures mining for graph
  recognition.
\newblock In P.~{Latorre Carmona}, J.~S. S{\'a}nchez, and A.~L.~N. Fred,
  editors, \emph{{Pattern Recognition - Applications and Methods}}, volume 204,
  pages 79--91. Springer Berlin - Heidelberg, Berling, Germany,
  2013{\natexlab{b}}.
\newblock ISBN 978-3-642-36529-4.
\newblock \doi{10.1007/978-3-642-36530-0_7}.

\bibitem[Livi et~al.(2013{\natexlab{c}})Livi, Tahayori, Sadeghian, and
  Rizzi]{t2apdiss__ifsanafips2013}
L.~Livi, H.~Tahayori, A.~Sadeghian, and A.~Rizzi.
\newblock Aggregating $\alpha$-planes for type-2 fuzzy set matching.
\newblock In \emph{Proceedings of the Joint IFSA World Congress and NAFIPS
  Annual Meeting}, pages 860--865, Edmonton, AB, Jun. 2013{\natexlab{c}}.
\newblock \doi{10.1109/IFSA-NAFIPS.2013.6608513}.

\bibitem[Livi et~al.(2014)Livi, Rizzi, and Sadeghian]{odse}
L.~Livi, A.~Rizzi, and A.~Sadeghian.
\newblock Optimized dissimilarity space embedding for labeled graphs.
\newblock \emph{Information Sciences}, 266:\penalty0 47--64, 2014.
\newblock ISSN 0020-0255.
\newblock \doi{10.1016/j.ins.2014.01.005}.

\bibitem[Livi et~al.(2015)Livi, Rizzi, and Sadeghian]{si_asoc_grc}
L.~Livi, A.~Rizzi, and A.~Sadeghian.
\newblock Granular modeling and computing approaches for intelligent analysis
  of non-geometric data.
\newblock \emph{Applied Soft Computing}, 27:\penalty0 567--574, 2015.
\newblock ISSN 1568-4946.
\newblock \doi{10.1016/j.asoc.2014.08.072}.

\bibitem[Lov{\'a}sz(1996)]{Lovasz1996}
L.~Lov{\'a}sz.
\newblock {Random Walks on Graphs: A Survey}.
\newblock In D.~{Mikl{\'o}s}, V.~T. {S{\'o}s}, and T.~{Sz\H{o}nyi}, editors,
  \emph{{Combinatorics, Paul Erd\H{o}s is Eighty}}, volume~2, pages 353--398.
  J{\'a}nos Bolyai Mathematical Society, Budapest, 1996.

\bibitem[Maiorino et~al.(2015)Maiorino, Livi, Giuliani, Sadeghian, and
  Rizzi]{mixbionets2}
E.~Maiorino, L.~Livi, A.~Giuliani, A.~Sadeghian, and A.~Rizzi.
\newblock Multifractal characterization of protein contact networks.
\newblock \emph{Physica A: Statistical Mechanics and its Applications},
  428:\penalty0 302--313, 2015.
\newblock ISSN 0378-4371.
\newblock \doi{10.1016/j.physa.2015.02.026}.

\bibitem[Niwa et~al.(2009)Niwa, Ying, Saito, Jin, Takada, Ueda, and
  Taguchi]{niwa2009}
T.~Niwa, B.-W. Ying, K.~Saito, W.~Jin, S.~Takada, T.~Ueda, and H.~Taguchi.
\newblock {Bimodal protein solubility distribution revealed by an aggregation
  analysis of the entire ensemble of Escherichia coli proteins}.
\newblock \emph{Proceedings of the National Academy of Sciences}, 106\penalty0
  (11):\penalty0 4201--4206, 2009.
\newblock \doi{10.1073/pnas.0811922106}.

\bibitem[Norris(1998)]{norris98Markovchains}
J.~R. Norris.
\newblock \emph{{Markov Chains}}.
\newblock Cambridge University Press, Cambridge, UK, 1998.
\newblock ISBN 9780521633963.

\bibitem[Pedrycz and Gomide(1998)]{pedrycz1998introduction}
W.~Pedrycz and F.~Gomide.
\newblock \emph{{An Introduction to Fuzzy Sets: Analysis and Design}}.
\newblock Mit Press, Cambridge, MA, 1998.
\newblock ISBN 9780262161718.

\bibitem[Ramshini et~al.(2011)Ramshini, Parrini, Relini, Zampagni, Mannini,
  Pesce, Saboury, Nemat-Gorgani, and Chiti]{ramshini2011large}
H.~Ramshini, C.~Parrini, A.~Relini, M.~Zampagni, B.~Mannini, A.~Pesce, A.~A.
  Saboury, M.~Nemat-Gorgani, and F.~Chiti.
\newblock {Large proteins have a great tendency to aggregate but a low
  propensity to form amyloid fibrils}.
\newblock \emph{PLoS ONE}, 6\penalty0 (1):\penalty0 e16075, 2011.

\bibitem[R{\'e}nyi(1961)]{rrnyi1961measures}
A.~R{\'e}nyi.
\newblock {On measures of entropy and information}.
\newblock In \emph{{Fourth Berkeley Symposium on Mathematical Statistics and
  Probability}}, pages 547--561, 1961.

\bibitem[Rizzi et~al.(2013{\natexlab{a}})Rizzi, Livi, Tahayori, and
  Sadeghian]{t2vsdiss__ifsanafips2013}
A.~Rizzi, L.~Livi, H.~Tahayori, and A.~Sadeghian.
\newblock Matching general type-2 fuzzy sets by comparing the vertical slices.
\newblock In \emph{Proceedings of the Joint IFSA World Congress and NAFIPS
  Annual Meeting}, pages 866--871, Edmonton, AB, Jun. 2013{\natexlab{a}}.
\newblock \doi{10.1109/IFSA-NAFIPS.2013.6608514}.

\bibitem[Rizzi et~al.(2013{\natexlab{b}})Rizzi, Possemato, Livi, Sebastiani,
  Giuliani, and {Frattale Mascioli}]{grapsec_ijcnn_2013}
A.~Rizzi, F.~Possemato, L.~Livi, A.~Sebastiani, A.~Giuliani, and F.~M.
  {Frattale Mascioli}.
\newblock A dissimilarity-based classifier for generalized sequences by a
  {G}ranular {C}omputing approach.
\newblock In \emph{{Proceedings of the 2013 International Joint Conference on
  Neural Networks}}, pages 2397--2404, Dallas, USA, Aug 2013{\natexlab{b}}.
\newblock ISBN 978-1-4673-6129-3.
\newblock \doi{10.1109/IJCNN.2013.6707041}.

\bibitem[Sch{\"o}lkopf and Smola(2002)]{schoelkopf+smola2002}
B.~Sch{\"o}lkopf and A.~J. Smola.
\newblock \emph{{Learning with Kernels: Support Vector Machines,
  Regularization, Optimization, and Beyond}}.
\newblock MIT Press, Cambridge, MA, 2002.
\newblock ISBN 9780262194754.

\bibitem[Taniuchi and Anfinsen(1969)]{Tanuichi}
H.~Taniuchi and C.~B. Anfinsen.
\newblock {An Experimental Approach to the Study of the Folding of
  Staphylococcal Nuclease}.
\newblock \emph{Journal of Biological Chemistry}, 244\penalty0 (14):\penalty0
  3864--3875, 1969.

\bibitem[Tasdighian et~al.(2013)Tasdighian, {Di Paola}, {De Ruvo}, Paci,
  Santoni, Palumbo, Mei, {Di Venere}, and Giuliani]{tasdighian2013modules}
S.~Tasdighian, L.~{Di Paola}, M.~{De Ruvo}, P.~Paci, D.~Santoni, P.~Palumbo,
  G.~Mei, A.~{Di Venere}, and A.~Giuliani.
\newblock {Modules identification in protein structures: the topological and
  geometrical solutions}.
\newblock \emph{Journal of Chemical Information and Modeling}, 54\penalty0
  (1):\penalty0 159--168, 2013.
\newblock \doi{10.1021/ci400218v}.

\bibitem[Thompson(2005)]{cca}
B.~Thompson.
\newblock \emph{{Canonical Correlation Analysis}}.
\newblock John Wiley \& Sons, Hoboken, NJ, 2005.
\newblock ISBN 9780470013199.

\bibitem[Waldo(2003)]{Waldo200333}
G.~S. Waldo.
\newblock {Genetic screens and directed evolution for protein solubility}.
\newblock \emph{Current Opinion in Chemical Biology}, 7\penalty0 (1):\penalty0
  33--38, 2003.
\newblock ISSN 1367-5931.
\newblock \doi{10.1016/S1367-5931(02)00017-0}.

\bibitem[Zbilut et~al.(2007)Zbilut, Chua, Krishnan, Bossa, Rother, Webber, and
  Giuliani]{PROT:PROT21179}
J.~P. Zbilut, G.~H. Chua, A.~Krishnan, C.~Bossa, K.~Rother, C.~L. Webber, and
  A.~Giuliani.
\newblock {A topologically related singularity suggests a maximum preferred
  size for protein domains}.
\newblock \emph{Proteins: Structure, Function, and Bioinformatics}, 66\penalty0
  (3):\penalty0 621--629, 2007.
\newblock ISSN 1097-0134.
\newblock \doi{10.1002/prot.21179}.

\end{thebibliography}
\end{document}